\documentclass[aps,pre, preprint,floatfix,11pt]{revtex4}
\usepackage{amssymb,epsf}

\usepackage{latexsym}
\usepackage{xcolor}
\usepackage{epsfig}
\usepackage{float} 
\usepackage{color} 
\usepackage{amsmath}
\usepackage{bm}
\usepackage{amsthm}
\usepackage{amssymb}
\usepackage{amssymb,epsf}
\usepackage{latexsym}
\usepackage{epsfig}
\usepackage{graphicx}

\begin{document}
\title{Type-III intermittency in emergent bursting dynamics of globally coupled rotators}
\author{Marzena Ciszak}
\affiliation{CNR-Istituto Nazionale di Ottica c/o LENS, I-50019 Sesto Fiorentino (FI), Italy.}
	\author{Francesco Marino}
	\affiliation{CNR-Istituto Nazionale di Ottica c/o LENS and INFN, Sezione di Firenze, I-50019 Sesto Fiorentino (FI), Italy.}

\begin{abstract}
Globally coupled populations of phase rotators with linear adaptive coupling can exhibit collective bursting oscillations between asynchronous and partially synchronized states, which can be either periodic or chaotic. Here, we analyze the transition between these two regimes, where the dynamics consists of periods of nearly regular bursting interspersed with irregular spiking intervals, and demonstrate its correspondence to intermittent transition to chaos. Specifically, we consider a bimodal Kuramoto model with linear global feedback, which allows for a mean-field formulation of the dynamics and thus to investigate the phenomenology in the thermodynamic limit. We reconstruct the one-dimensional first-return maps of inter-burst intervals and estimate the Floquet multiplier associated with the unstable bursting solution. The results indicate type-III intermittency, which is also supported by the scaling of the average laminar periods as the control parameter varies, along with their probability density distribution.
\end{abstract}

\maketitle

\section{introduction}

Over the past few decades, complex networks of phase rotators, often described using the Kuramoto model \cite{kuramoto} and its extensions \cite{acebron2005}, have been widely studied as models for the emergence of non-trivial collective dynamics. These phenomena include quasi-periodicity, period-doubling cascades to chaos and attractor crises (see e.g. \cite{matthews1990, hakim1992, nakagawa1993, so2011}), as well as first-order phase transitions between states with varying degrees of synchronization \cite{tanaka1997, pazo2009, olmi2014}. A particularly intriguing research direction focuses on the emergence of collective slow-fast phenomena, which are characteristic of low-dimensional dynamical systems where two or more variables evolve on very different time scales. The separation between time scales produces oscillatory patterns with characteristic square-wave-like shapes, consisting of slowly varying periods of evolution around two amplitude levels interrupted by fast transitions between them. The slow dynamics occurs near the attracting branches of a critical manifold the fast transitions are trajectories transeversal to these branches \cite{smale}. Systems exhibiting these dynamics over a range of parameters also display excitability, meaning they can return to a stable equilibrium after a perturbation through a phase space excursion that is independent of the applied stimulus.

If a population of rotators exhibits a hysteretic phase transition between asynchronous and partially synchronized states, a relatively simple self-sustained adaptation mechanism can induce collective slow-fast dynamics \cite{skardal2014,noi}. The adaptive linear feedback drives the system across the phase transition on a slow time scale, which can result in collective excitability, periodic and chaotic bursting \cite{noi}, as well as canard explosions \cite{coll-can}. While these macroscopic phenomena are fairly reminiscent of the slow-fast dynamics observed in low-dimensional neuron models, like the Hindmarsh-Rose \cite{hr, wang, gmiranda, innocenti2007} and the FitzHugh-Nagumo model with inertia (FHNI model) \cite{marino2007}, in the above networks they gradually emerge as the size of the population increases, additionally showing a remarkable robustness against network diluteness \cite{paolini}. Collective slow-fast dynamics, in the form of mixed-mode oscillations \cite{dolcemascolo} and canard cascading \cite{otti,balzer}, have been also recently reported in networks with adaptive mean-field coupling. However, in those studies, the emergence of these dynamics depends on the slow-fast nature of the nodes, whereas here, as in Refs. \cite{skardal2014,noi,coll-can,paolini}, the populations consist of simple phase rotators.

In the case of populations with a bimodal Lorentzian distribution of natural frequencies, an exact three-dimensional mean-field formulation can be derived, which allows to reconstruct the complete bifurcation diagram and analyse the critical manifold which determines the dynamics \cite{noi,coll-can}.
However, many important features related to these phenomena, particularly the mechanisms underlying the transitional regime from periodic to chaotic collective bursting, remain to be clarified. In this regime, the time series exhibit laminar sequences of nearly regular bursts interspersed with irregular (chaotic) periods. Within the framework of the FHNI model, the average residence times in the nearly-periodic firing state have been found to follow the same scaling law as that observed at the onset of type I intermittency, characteristic of systems near a saddle-node bifurcation \cite{ciszak23}. 

In this work, we show that an intermittent transition to chaos also underlies the collective bursting dynamics of large populations of rotators subject to adaptive coupling, though with characteristics indicative of type-III intermittency. 

The paper is organized as follows. In Sec. II, we introduce the network model and its mean-field description, consisting of a 3D slow-fast system with two fast and one slow variable. In Sec. III we provide evidence of intermittency in the network dynamics, and in Sec. IV, we quantitatively analyze the transition in the thermodynamic limit using an exact mean-field model. Specifically, we reconstruct the one-dimensional first-return maps of inter-burst intervals showing that, at the transition, a Floquet multiplier associated with the periodic bursting solution crosses the unit circle at the real value of $-1$, that uniquely identifies intermittency as type III \cite{pomeau, schuster}. We support this result by determining the critical exponents that govern the scaling of the average duration and the probability density of laminar phases. Additionally, we identify other characteristic features of this dynamics, such as an oscillatory growth in the inter-burst intervals, linked to the onset of a period doubling mode within the laminar region. Finally, we briefly discuss these results within the framework of a geometric singular perturbation analysis of the mean-field dynamics. Conclusions and future perspectives are presented in Sec. V.

\section{Network model and mean-field formulation}
\label{models}

We consider a globally-coupled population of $N$ phase rotators with adaptive coupling strength $S(t)$ \cite{noi}
\begin{subequations}
\label{network} 
\begin{eqnarray}
{\dot \theta}_k(t) &=& \omega_k + \frac{S(t)}{N} \sum_{j=1}^N \sin(\theta_j(t)- \theta_k(t)), \;  \qquad k=1,\ldots, N \label{net1} \\
{\dot S}(t) &=& \epsilon \left[-S(t) + K -\alpha R(t) \right] \;  \label{net2}
\end{eqnarray}
\end{subequations}
where $\theta_k$ are the phases of each oscillator and $\omega_k$ their natural frequencies. The coupling strength in Eq. (\ref{net1}) is replaced by a dynamic variable $S(t)$, which is governed by the level of synchronization between the oscillators. This is quantified by the variable $R$ ($0 < R < 1$), representing the modulus of the complex Kuramoto order parameter $Z(t) = \frac{1}{N} \sum_{j=1}^N {\rm e}^{i \theta_j(t)} = R(t) {\rm e}^{i \phi(t)}$ \cite{kuramoto}. 

We consider a bimodal distribution of natural frequencies, represented as the sum of two Lorentzian distributions centered at $\pm \omega_0$, each with a half-width at half-maximum of $\Delta$.  
The distribution is generated by applying the following deterministic rule  \cite{montbrio2015}
\begin{subequations}
\label{gend} 
\begin{eqnarray}
\omega_j &=& -\omega_0 + \Delta \tan (\pi\xi_j/2)  \qquad \xi_j = \frac{2j-N/2-1}{N/2+1} \qquad j=1 \dots N/2\\
\omega_j &=& \omega_0 + \Delta \tan (\pi\xi_j/2) \qquad \xi_j = \frac{2j-3N/2-1}{N/2+1} \qquad j=N/2+1 \dots N
\end{eqnarray}
\end{subequations}  
The deterministically-generated frequency distribution allows for a reduction of finite-size effects with respect to randomly-generated distributions and faster convergence to mean-field dynamics \cite{note}. Remarkably, even with as few as $N = 1000$ oscillators, the network time-series exhibits the characteristic bursting shape, closely matching the results produced by the mean-field model \cite{coll-can}.

We fix the distribution parameters at $\omega_0 = 1.8$ and $\Delta = 1.4$, values for which, in the absence of feedback ($\alpha = 0$), the system exhibits a first-order hysteretic transition from incoherent to partially synchronized states \cite{martens2009, pazo2009}. The feedback loop gain is set to $\alpha = 5$, while $\epsilon = 0.01$ represents the ratio between the characteristic time scale of the macroscopic network dynamics and that of the feedback. Since $\epsilon \ll 1$, the coupling adapts slowly to the rapid switching between incoherent and coherent states, driving the system across the hysteretic transition. It is within this regime that the competition between the macroscopic network dynamics and feedback, operating on different time scales, leads to collective slow-fast phenomena.

Although these phenomena do not rely on a specific shape of the frequency distribution (as long as a hysteretic phase transition is present) \cite{noi}, a bimodal distribution enables the derivation of an exact mean-field formulation of the problem following the approach in Refs. \cite{martens2009,so2011} based on the the Ott-Antonsen Ansatz \cite{ott2008}. 
Decomposing the Kuramoto paramter as $Z = \frac{1}{2}(z_1 + z_2)$ 
where $z_k=\rho_k{\rm e}^{i \phi_k}$ ($k=1,2$) are the complex order parameters corresponding to the oscillators associated with each Lorentzian distribution, and assuming $\rho_1 \approx \rho_2 = \rho$ we obtain \cite{noi}
\begin{subequations}
\label{mf} 
\begin{eqnarray}
\dot{\rho} & = & -\Delta \rho + \frac{S}{4} \rho (1 - \rho^2) (1 + \cos(\phi)) \;  \label{eq:model1} \\
\dot{\phi} & = & 2 \omega_0 - \frac{S}{2} (1 + \rho^2) \sin(\phi) \;  \label{eq:model2} \\
\dot{S} & = & -\epsilon \left[ S - K + \alpha \rho \sqrt{\frac{1 + \cos(\phi)}{2}} \right] \;  \label{eq:model3} 
\end{eqnarray}
\end{subequations}
where $\phi = \phi_2 - \phi_1$, $\pm \omega_0$ are the centers of the two Lorentzian peaks of the distribution, $\Delta$ is their half-width at half-maximum and $R = \rho \sqrt{(1 + \cos(\phi))/2}$. 

\section{Intermittent network dynamics}
\begin{figure*}
  %  \centering
    \includegraphics*[width=1.\columnwidth]{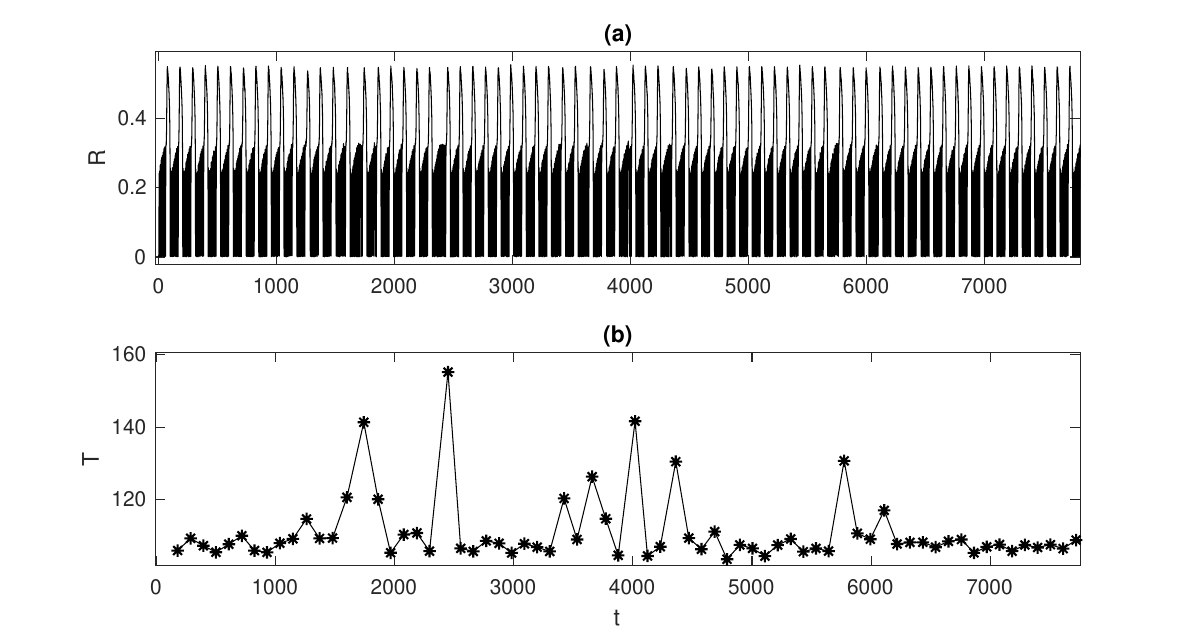}
    \caption{(a) Time-series of $R$ for the network model (\ref{network}) for $N=2 \times 10^5$ rotators and $K=7.42$. (b) Time-series of the corresponding inter-burst intervals $T$. 
    Other parameters as in the text.}
    \label{fig:1n}
\end{figure*}

We now briefly summarize the dynamical regimes observed from integrating the network model (\ref{network}). In the absence of adaptation and for the parameters specified above, the network exhibits a hysteretic first-order transition between asynchronous and partially synchronized states \cite{pazo2009}. When feedback is introduced, the system alternates between partially synchronized phases, characterized by bursts with well-defined shape, amplitude, and duration, and desynchronized phases marked by oscillatory values of $R$ (spikes). Depending on the value of the control parameter $K$, the number of spikes between to subsequent bursts can be always the same, or become irregular, leading to a chaotic bursting phase. These regimes have been extensively studied in Refs. \cite{noi,coll-can}.

\begin{figure}
    \centering
    \includegraphics[width=1.\linewidth]{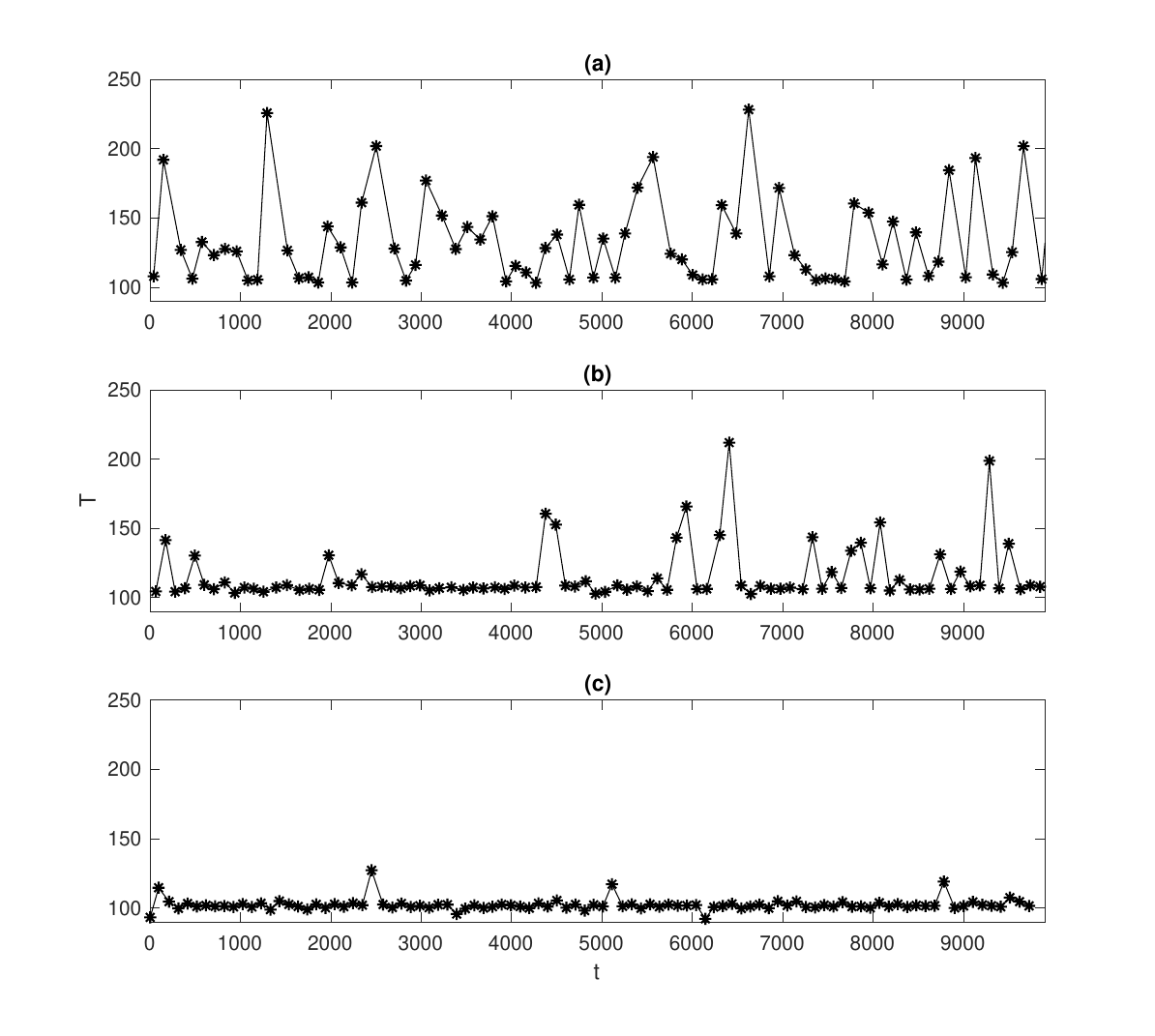}
    \caption{Evolution of $T$ for the network model (\ref{network}) for $N=2 \times 10^5$ rotators: (a) $K=7.41$ (b) $K=7.42$. and (c) $K=7.45$. Other parameters as in the text.}
    \label{fig:2n}
\end{figure} 

Here, we focus on the transition between these regimes, where long laminar phases of periodic bursting are intermittently interrupted by short periods of irregular spiking. An example of this behavior is shown in Fig. \ref{fig:1n} (a), which presents the time series of $R$ obtained through numerical integration of Eq. (\ref{network}). Intervals of irregular dynamics manifest as elongated periods between adjacent bursts, as observed, for instance, around $t = 1700$ and $t = 2400$. These irregular phases are better visualized by plotting the evolution of the successive inter-burst intervals $T$ (see  Fig. \ref{fig:1n} (b)), where it is clear that laminar periods identified by a mean inter-burst interval $\overline{T_l} \approx 113.82 $ are occasionally interrupted by irregular spiking events, causing an increase in $T$. At lower values of $K$, for which the system fully enters the chaotic bursting regime, the laminar phases become very short and rare (see Fig. \ref{fig:2n} (a)). On the other hand, the duration and frequency of irregular periods decreases as the network approaches periodic bursting (see Fig. \ref{fig:2n} (b-c)).
Although this behavior clearly points to a kind of intermittency, the finite size fluctuations associated with $R \sim \mathcal{O}(1/\sqrt{N})$, prevent the quantitative analysis of the phenomenon, in spite of the large number $N$ of oscillators involved ($N=2 \times 10^5$) and the use of the deterministically-generated frequency distribution (\ref{gend}). Moreover, noise fluctuations generally have a significant impact on the scaling laws that characterize intermittency \cite{noiseInterm_a, noiseInterm_b} even in low-dimensional systems, and tend to smooth out the hysteretic transitions that, in our network, govern the dynamics. In particular, in bursting oscillators they strongly affect the burst orbits \cite{incoherence}, which would lead to a modification of the natural period of the laminar phase. In the next section, we will thus perform a quantitative analysis of the phenomenon in the thermodynamic limit by means of the mean-field model (\ref{eq:model3}). 

\section{Analysis of intermittency in the thermodynamic limit}

\begin{figure*}
  %  \centering
    \includegraphics*[width=1.\columnwidth]{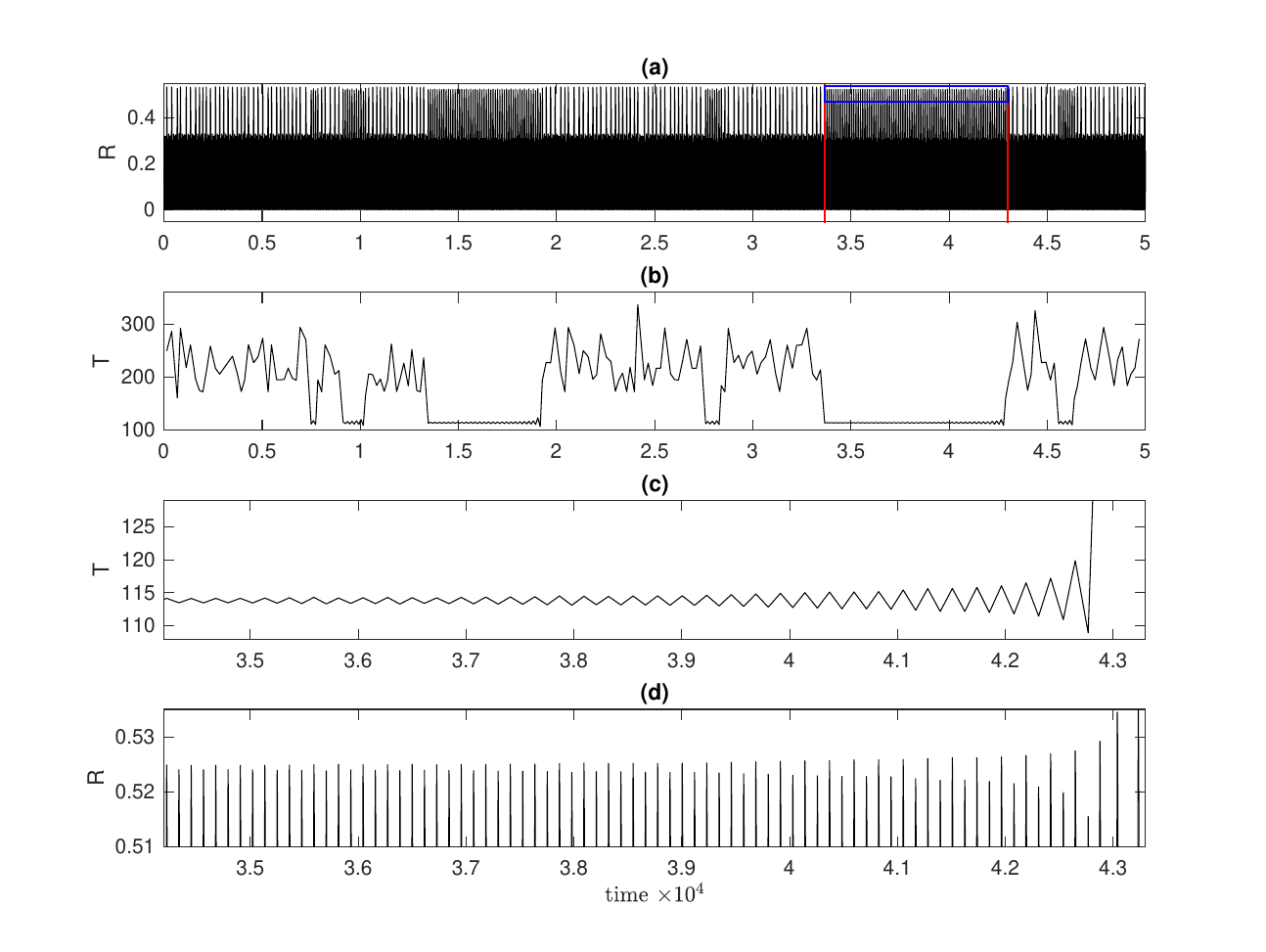}
    \caption{Evolution of the mean-field model (\ref{mf}) for $K=7.40938789$. (a) Time series for a Kuramoto order parameter $R$, and (b) corresponding successive $T$ in time. c) Blow up of successive $T$
    in the laminar phase marked by the red vertical lines in panel a). d) Zoom-in of the maxima of R within the same laminar phase (see blue box in panel (a)).}
    \label{fig:3n}
\end{figure*}

In Fig. \ref{fig:3n}(a) we plot the typical time trace for $R$ in the intermittent regime, obtained by numerical integration of Eq. (\ref{eq:model3}). The time series show the usual intermittent behaviour with alternation between laminar and aperiodic bursting periods. As observed in the network (see Figs. \ref{fig:1n}(b) and \ref{fig:2n}), the intermittent dynamics is highlighted by plotting the time evolution of the corresponding inter-burst periods $T$, as shown in  Fig. \ref{fig:3n}(b). Of particular interest, is the evolution of $T$ within a laminar region, a detailed view of which is shown in the magnified section in Fig. \ref{fig:3n}(c). An oscillatory growth of $T$ is clearly visible, reflecting a similar oscillatory evolution of the burst maxima (see, Fig. \ref{fig:3n}(d)). A closer inspection of these maxima also reveal the growth of a period-doubled amplitude together with a decrease of the fundamental amplitude. When this subharmonic component reaches a certain maximum value, the system enters the chaotic dynamics. 

An inter-burst interval oscillating with increasing amplitude in the vicinity of an irregular phase is a characteristic feature of type-III intermittency \cite{schuster}, in contrast to the monotonic or spiraling patterns seen in type I and type II, respectively. Moreover, type-III intermittency arises from a subcritical period-doubling bifurcation, where an unstable period-2 orbit collides with and destabilizes a stable period-1 orbit \cite{dubois,ott}. Consequently, the time series are typically characterized by the gradual appearance of a period-2 component at the end of a laminar period, on the line of what shown in Fig. \ref{fig:3n} (d).

When a laminar phase of periodic bursting is sporadically interrupted by periods of irregular spiking, it indicates the instability of the periodic limit cycle. The instability can be characterized by means of one-dimensional first-return maps of inter-burst intervals and is signaled when the modulus of at least one Floquet multiplier exceeds one. Intermittency is classified into types I, II, and III depending on how the Floquet multiplier crosses the unit circle in the complex plane \cite{pomeau,ott}. In type-I intermittency, a real multiplier crosses the unit circle at $+1$, whereas two complex conjugate multipliers crossing the unit circle simultaneously indicate intermittency of type-II. Type-III intermittency is instead identified by a real Floquet multiplier that crosses the unit circle at $-1$. 

\begin{figure}
    \centering
    \includegraphics[width=1.\linewidth]{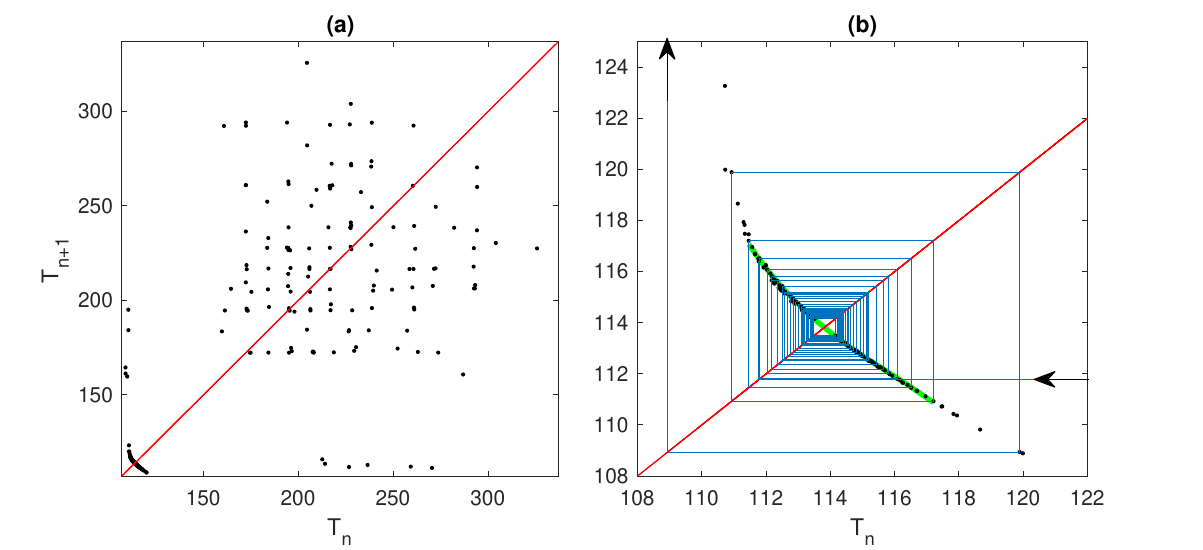}
    \caption{(a) The first return map for $T_n$ from Fig. \ref{fig:3n}(b). (b) The first return map for $T_n$ within the selected time interval shown in Fig. \ref{fig:3n}(c). Green line plots $x_n+\overline{T_l}$, red line is a diagonal and blue line marks the successive iteration steps. Black arrows show the flow directions.}
    \label{fig:4n}
\end{figure}

The return map linking consecutive inter-burst intervals is displayed in Fig. $\ref{fig:4n} (a)$. For low inter-burst intervals, corresponding to the laminar phase of nearly-periodic bursting, the motion remains confined into a small region of the plane for a large number of iterations. The local map in this region determines the intermittency type, while a reinjection process drives the trajectories back from the chaotic phase of large $T_n$. A zoom-in of the region with low $T_n$ values (see Fig. \ref{fig:4n} (b)) reveals the oscillatory growth of inter-burst intervals, a hallmark of the laminar phase in type-III intermittency. 

We fit the curve formed by the iteration points $x_n=T_n- \overline{T_l}$, where $\overline{T_l}$ is the mean inter-burst interval during the laminar phases, using a polynomial cubic function
\begin{eqnarray}
x_{n+1}=f(x_n)=x_n(ax_n^2+bx_n+c)
\label{map}
\end{eqnarray}
and obtain the coefficients $a=-0.01852$, $b=0.1105$ and $c=-1.025$ with residuals $<5\%$. The real fixed point is $x^*=0$, and the corresponding real eigenvalue is
\begin{eqnarray}
\lambda=f'(x^*)=-1.025 \, ,
\end{eqnarray}
which indicates type-III intermittency.

The reinjection mechanism is determined by the stability properties of a one-dimensional critical manifold that organizes the collective bursting dynamics on a slow timescale. A geometric singular perturbation analysis of the model (\ref{mf}) has been presented in Refs. \cite{noi,coll-can}. Here, we just comment on specific features that are relevant for intermittency.

We recall that on the slow time scale corresponding to the duration of a burst, $\tau = \varepsilon t$, the system evolves according to the feedback equation (\ref{eq:model3}) while satisfying the algebraic constraint $(\dot{\rho},\dot{\phi})$=($0,0$). The fixed points of the fast subsystem described by Eqs. (\ref{eq:model1}-\ref{eq:model2}) thus define the critical manifold on which the slow dynamics occurs. These points indeed lay on the curve $\Sigma= \Sigma_0 \cup \Sigma_{\rho}$, where $\Sigma_0$ 
is given by the set of incoherent steady-state solutions $\Sigma_0 = \{\rho_s = 0, \sin(\phi_s) = 4 \omega_0 /S, S\}$, and $\Sigma_{\rho}= \{\rho_s, \phi_s, S\}$ 
is defined by the equations $S (1+\rho_s) \sin(\phi_s)=4 \omega_0$ and 
\begin{equation}
S =\frac{2\omega_0^2}{\Delta} \frac{1-\rho^2}{(1+\rho^2)^2} + \frac{2 \Delta }{1-\rho^2} \equiv \mathcal{F}(\rho) \, .
\label{eq_cm}
\end{equation}
Using the stationary solution of Eq. (\ref{eq:model1}) and the relation $R = \rho \sqrt{(1 + \cos(\phi))/2}$, Eq. (\ref{eq_cm}) can be expressed in terms of $R$, as
\begin{equation}
\label{nm}
S =(S R^2 + 2 \Delta)\left(1+ \frac{\omega_0^2}{S R^2 + \Delta} \right) \, .
\end{equation}

\begin{figure}
    \centering
    \includegraphics[width=1.\linewidth]{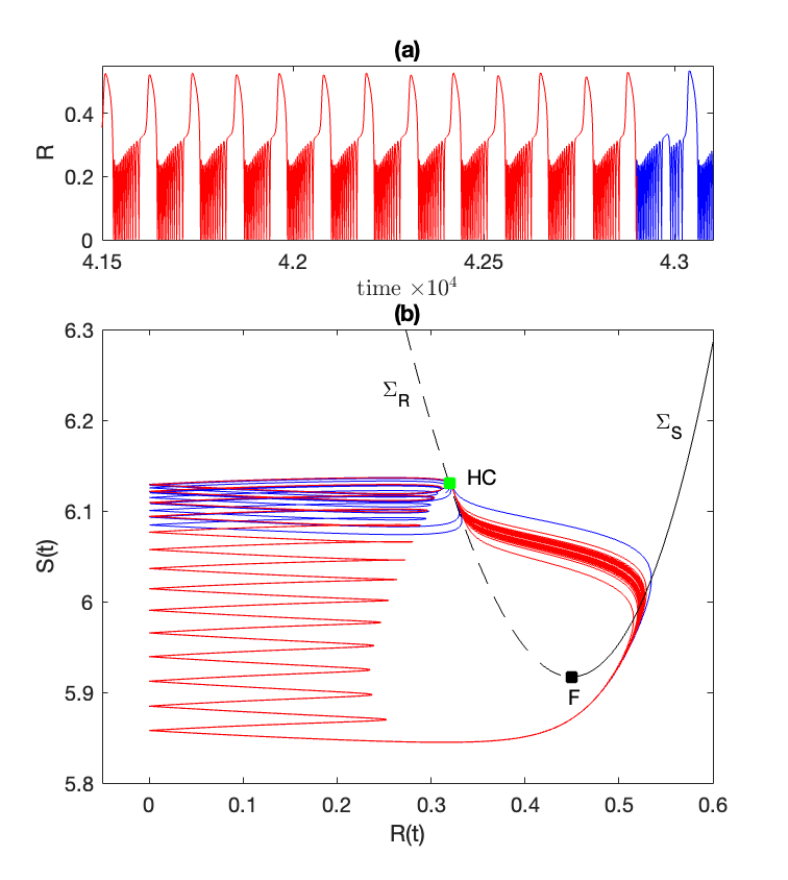}
    \caption{ (a) Time series for $R$ obtained from the mean-field model with $K=7.4093878$. Red color marks nearly regular spiking within the laminar phase while the blue color marks the abrupt emergence of the chaotic spiking. (b) Projection on the $(R,S)$ plane disclosing the bursting regime: solid and dashed black lines indicate the attracting $\Sigma _S$ and repelling $\Sigma _R$ manifolds, respectively. The green dot marks the injection point $HC$ into the chaotic phase and the black dot marks the fold point $F$.}
    \label{fig:5n}
\end{figure}

By linearizing the fast subsystem along the aforementioned curve, we observe that it comprises a branch of stable equilibria $\Sigma_S$ (represented by the solid line in Fig. \ref{fig:5n}) and an unstable branch $\Sigma_R$ (dashed line), which coalesce in a saddle-node bifurcation at the fold point $F$. For $\omega_0 > \Delta$, equilibria along $\Sigma_0$ are always unstable; however, a stable limit cycle solution emerges from a degenerate supercritical Hopf bifurcation. 
Since the trajectories Eqs. (\ref{mf}) are attracted by stable branches of $\Sigma$ while will be repelled by the unstable ones \cite{Fenichel}, this serves as the underlying mechanism for the observed intermittency, as we explain below.

In Fig. \ref{fig:5n}(a) we plot a numerical time-series of $R$ in the intermittent bursting regime, showing laminar phase (red trace) suddenly interrupted by an irregular spiking interval (blue trace). The same trajectory is plotted Fig. \ref{fig:5n}(b) on the $(R, S)$ plane, along with the projection of the attracting and repelling branches $\Sigma_S$ and $\Sigma_R$ given by Eq. (\ref{nm}). 

Starting, for example, from a point near $\Sigma_S$ (solid black line) the trajectory monotonically evolves towards the fold point $F$. At this point, the system leaves the critical manifold and moves into a region where the fast subsystem exhibits the aforementioned stable limit cycle. In this region, the previous monotonic evolution transforms into a sequence of periodic spikes. 
This oscillatory trajectories continues until it collides with the repelling branch of the manifold, where it disappears through a saddle homoclinic-orbit bifurcation (green point $HC$ in Fig. \ref{fig:5n}(b)). Since $\Sigma_{R}$ repels all neighboring trajectories while $\Sigma_{S}$ attract them, the motion is driven back to the upper state where a new burst orbits begins.
nterestingly, this transition displays the characteristic features of canard explosions \cite{canard1, canard2}. Due to the finite value of$\varepsilon$, the trajectories do not immediately depart from the region near the repelling part of the manifold; instead, they can slowly flow close to it for a significant amount of time.
Since canard orbits lye at the intersection of attracting and repelling slow manifolds, they exhibit extreme sensitivity to fluctuations, as evidenced by the spread of trajectories observed in the vicinity of $\Sigma_R$. Such small variations in the burst orbits hamper the homoclinic connection, causing some trajectories to be repelled back to the spiking region, breaking the periodicity.

The above scenario, in which transition from the upper to lower spiking state occurs through the fold point $F$ and the spiking phase disappears via $HC$ bifurcation is known as \emph{fold-homoclinic} bursting \cite{izhikevich}, and it is found in several models of neuronal activity, including e.g. the Hindmarsh-Rose \cite{hr} and the Morris-–Lecar system with current-feedback control \cite{izhikevich}.
 
\begin{figure}
    \centering
 \includegraphics[width=1.\linewidth]{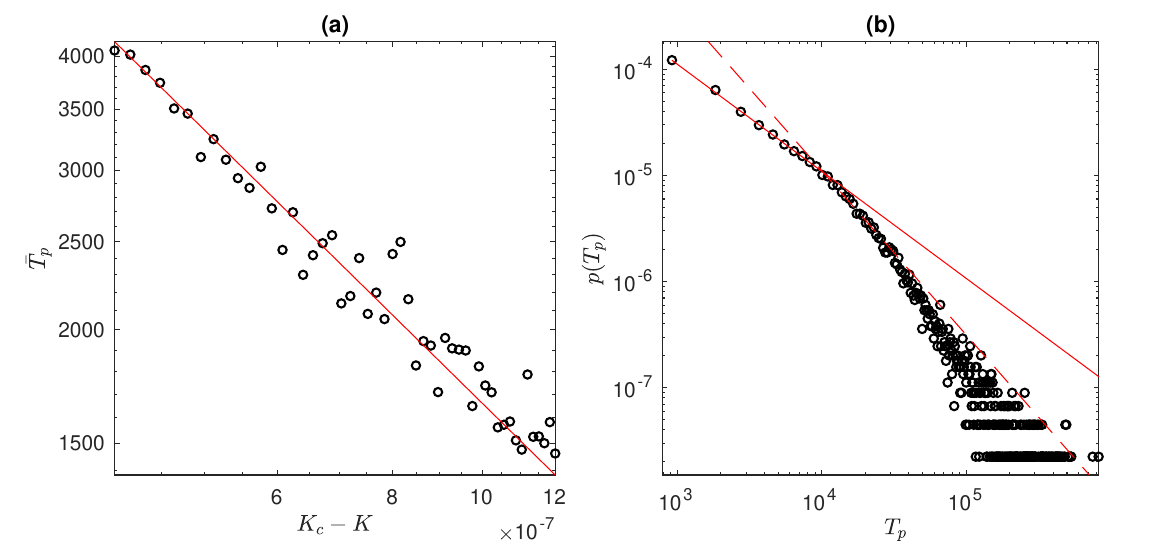}
    \caption{Statistical analysis of the mean-field intermittent dynamics. (a) The average laminar length as control parameter $K$ approaches its critical value $K_c$. The red solid line shows $g(K)=(K_c-K)^{\gamma}$ with $\gamma=-0.9317 \pm 0.0519$ and $K_c=7.4093882$. (b) Probability density of the laminar lengths for $K=7.40938789$. Lines depict $p(T_p)=T_p^{\beta _j}$ for $j=1,2$, where $\beta_1=-1.009\pm 0.026$  (red solid) and $\beta_2=-1.548\pm 0.1$ (red dashed). }
    \label{fig:6n}
\end{figure}

We finally analyse the mean length of the laminar phases $\overline{T_p}$ as the control parameter $K$ is varied near the transition from the chaotic to periodic regime (see Fig. \ref{fig:6n} (a)). Linear fitting of the data in a log-log scale gives a slope $\gamma=-0.9317 \pm 0.0519$, consistent with the power-law behavior $g(K) \sim (K_c-K)^\gamma$ with $\gamma=-1$, as predicted for type-III intermittency \cite{ott}. This would exclude type-I intermittency for which $\gamma=-1/2$, while would be still compatible with what expected for intermittency of type II.
A discriminant is given by the probability density distribution of the laminar phases $p(T_p)$. The theory of type-III intermittency predicts $p(T_p) \sim T_p^{\beta}$ with $\beta=-3/2$. Such scaling is generally valid only for sufficiently large values of $T_p$ \cite{berge}, i.e. laminar phases involving a sufficiently high number of bursts.

An example of the probability density distribution is shown in Fig. \ref{fig:6n}(b). We observe two distinct scaling regions: one for short laminar phases with an exponent $\beta_1=-1.009\pm 0.026$ (solid line), and another for longer laminar phases with an exponent $\beta_2=-1.548\pm 0.1$ (dashed line). The latter is consistent with the$-3/2$ power-law scaling expected for longer values of $T_p$ . On the other hand, the approximate $1/T_p$ scaling for shorter laminar phases, to our knowledge, has not been previously reported in the intermittency literature and might be specific of the (slow-fast) dynamics of our system. This is an interesting aspect that deserves further investigation.

\section{Conclusions}
\label{disc_1}

We demonstrated intermittent behaviour in the collective bursting dynamics of a population of globally coupled phase-rotators. For specific values of the control parameter, phases of regular bursting are interrupted by irregular periods. By analyzing the network dynamics through an exact mean-field model, we can accurately reconstruct the return map of inter-burst intervals and extract the scaling laws for the average duration and frequency of laminar phases, effectively removing the influence of finite-size fluctuations. The results of our analysis are consistent with type-III intermittency. Our study demonstrates that a macroscopic intermittent dynamics can spontaneously arise in adaptive networks of rotators and that the phenomenon persists in the thermodynamic limit, ruling out any relation to finite-size effects. The reported dynamical macroscopic scenario are pretty reminiscent of that observed in low-dimensional neuron models, suggesting potential implications for computational neuroscience. The concept of intermittent synchronization \cite{yang,zhao,choudhary,vera} seems in fact an important mechanism in neural dynamics \cite{ahn}. It has been suggested to play a role in neuro-motor diseases such as Parkinson’s disease \cite{park,rat} or epilepsy \cite{koro} and has been also found in simple network models of bistable perception \cite{polonica}. More broadly, we expect our findings to offer insights into the mechanisms through which microscopic interactions can collectively lead to the spontaneous emergence of global intermittency, a process that may be relevant in the onset of certain turbulent dynamics.


\begin{thebibliography}{00}
\bibitem{kuramoto} Y. Kuramoto, "Self-entrainment of a population of coupled non-linear oscillators," International Symposium on Mathematical Problems in Theoretical Physics (Springer, Berlin, 1975), pp. 420–422.
\bibitem{acebron2005} J. A. Acebr\'on, L. L. Bonilla, C. J. P. Vicente, F. Ritort, and R. Spigler, "The Kuramoto model: A simple paradigm for synchronization phenomena," Rev. Mod. Phys. {\bf 77}, 137 (2005).
\bibitem{matthews1990} P. C. Matthews and S. H. Strogatz, "Phase diagram for the collective behavior of limit-cycle oscillators," Phys. Rev. Lett. {\bf 65}, 1701 (1990).
\bibitem{hakim1992} V. Hakim and W.-J. Rappel, "Dynamics of the globally coupled complex Ginzburg-Landau equation," Phys. Rev. A {\bf 46}, R7347 (1992).
\bibitem{nakagawa1993} N. Nakagawa and Y. Kuramoto, "Collective Chaos in a Population of Globally Coupled Oscillators," Prog. Theor. Phys. {\bf 89}, 313 (1993).
\bibitem{so2011} P. So and E. Barreto, "Generating macroscopic chaos in a network of globally coupled phase oscillators," Chaos {\bf 21}, 033127 (2011).
\bibitem{tanaka1997} H.-A. Tanaka, A. J. Lichtenberg, and S. Oishi, "First Order Phase Transition Resulting from Finite Inertia in Coupled Oscillator Systems," Phys. Rev. Lett. {\bf 78}, 2104 (1997).
\bibitem{pazo2009} D. Paz\'o and E. Montbri\'o, "Existence of hysteresis in the Kuramoto model with bimodal frequency distributions," Phys. Rev. E {\bf 80}, 046215 (2009).
\bibitem{olmi2014} S. Olmi, A. Navas, S. Boccaletti, and A. Torcini, "Hysteretic transitions in the Kuramoto model with inertia," Phys. Rev. E {\bf 90}, 042905 (2014).
\bibitem{smale}  M. W. Hirsch, R. L. Devaney, and S. Smale {\it Differential Equations, Dynamical Systems, and Linear Algebra, Vol.60} (Academic Press, New York, 1974).
\bibitem{skardal2014} P. S. Skardal, D. Taylor, and J.G. Restrepo, "Complex macroscopic behavior in systems of phase oscillators with adaptive coupling," Physica D {\bf 267}, 27(2014).
\bibitem{noi} M. Ciszak, F. Marino, A. Torcini and S. Olmi, "Emergent excitability in populations of nonexcitable units," Phys. Rev. E, {\bf 102}, 050201 (2020).
\bibitem{coll-can} M. Ciszak, S. Olmi, G. Innocenti, A. Torcini, and F. Marino, "Collective canard explosions of globally-coupled rotators with adaptive coupling," Chaos, Solitons Fractals {\bf 153}, 111592 (2021).
\bibitem{hr} J. L. Hindmarsh and R. Rose, "A model of neuronal bursting using three coupled first order differential equations," Proc. R. Soc. London, Ser. B. Biol. Sci. {\bf 221}, 87 (1984).
\bibitem{wang} X.-J. Wang, "Genesis of bursting oscillations in the Hindmarsh-Rose model and homoclinicity to a chaotic saddle," Physica D {\bf 62}, 263 (1993).
\bibitem{gmiranda} J. M. Gonzalez-Miranda, "Observation of a continuous interior crisis in the Hindmarsh–Rose neuron model," Chaos {\bf 13}, 845 (2003).
\bibitem{innocenti2007} G. Innocenti, A. Morelli, R. Genesio, and A. Torcini, "Dynamical phases of the Hindmarsh-Rose neuronal model," Chaos {\bf 17}, 043128 (2007).
\bibitem{marino2007} F. Marino, F. Marin, S. Balle, and O. Piro, "Chaotically Spiking Canards in an Excitable System with 2D Inertial Fast Manifolds," Phys. Rev. Lett. {\bf 98},074104 (2007).
\bibitem{paolini} G. Paolini, M. Ciszak, F. Marino, S. Olmi, and A. Torcini, "Collective excitability in highly diluted random networks of oscillators," Chaos {\bf 32}, 103108 (2022).
\bibitem{dolcemascolo} A. Dolcemascolo, A. Miazek, R. Veltz, F. Marino, and S. Barland, "Effective low-dimensional dynamics of a mean-field coupled network of slow-fast spiking lasers, Phys. Rev. E {\bf 101}, 052208 (2020).
\bibitem{otti} O. D’Huys, R. Veltz, A. Dolcemascolo, F. Marino, and S. Barland, "Canard resonance: on noise-induced ordering of trajectories in heterogeneous networks of slow-fast systems", Journal of Physics: Photonics {\bf 3}, 024010 (2021).
\bibitem{balzer} J. Balzer, R. Berner, K. L\"udge, S. Wieczorek, J. Kurths and Serhiy Yanchuk, "Canard cascading in networks with adaptive mean-field coupling"  	arXiv:2407.20758 (2024).
\bibitem{ciszak23} M. Ciszak, S. Balle, O. Piro, and F. Marino, "Intermittent chaotic spiking in the van der Pol FitzHugh-Nagumo system with inertia," Chaos, Solitons Fractals {\bf 167}, 113053 (2023).
\bibitem{pomeau} Y. Pomeau and P. Manneville, "Intermittent transition to turbulence in dissipative dynamical systems," Communications in Mathematical Physics, {\bf 74}(2), 189–197 (1980).
\bibitem{schuster} H. G. Schuster {\it Deterministic Chaos: An Introduction} (VCH, 1988).
\bibitem{montbrio2015} E. Montbri\'o, D. Paz\'o and A. Roxin, "Macroscopic Description for Networks of Spiking Neurons," Phys. Rev. X {\bf 5}, 021028 (2015).
\bibitem{note} As an example of this convergence, we observed that the accuracy achieved by $N=10^4$ rotators using the deterministic rule \ref{gend} is comparable to that obtained by $N=10^6$ rotators using randomly-generated natural frequencies.
\bibitem{martens2009} E. A. Martens, E. Barreto, S. H. Strogatz, E. Ott, P. So, and T. M. Antonsen, "Exact results for the Kuramoto model with a bimodal frequency distribution," Phys. Rev. E {\bf 79}, 026204 (2009).
\bibitem{ott2008} E. Ott and T.M. Antonsen, "Low dimensional behavior of large systems of globally coupled oscillators," Chaos {\bf 18}, 037113 (2008).
\bibitem{noiseInterm_a} J.-P. Eckmann, L. Thomas, and P. Wittwer, "Intermittency in the presence of noise," J. Phys. A: Math. Gen. {\bf 14}, 3153-3168 (1981).
\bibitem{noiseInterm_b} J.E. Hirsch, M. Nauenberg and D.J. Scalapino, "Intermittency in the presence of noise: A renormalization group formulation," Physics Letters A {\bf 87}, 391-393 (1982).
\bibitem{incoherence} M. Ciszak, "Stochastic incoherence in the response of rebound bursters," Phys. A: Stat. Mech. Appl. {\bf 389}, 2351 (2010).
\bibitem{dubois} M. Dubois, M. A. Rubio and P. Berge, "Experimental Evidence of Intermittencies Associated with a Subharmonic Bifurcation," Phys. Rev. Lett. {\bf 51}, 1446 (1983).
\bibitem{ott} E. Ott, {\it Chaos in Dynamical Systems} (Cambridge University Press, 1993).
\bibitem{Fenichel} N. Fenichel, "Geometric singular perturbation theory for ordinary differential equations," J. Diff. Eq. {\bf 31}, 53 (1979).
\bibitem{canard1} J. L. Callot, F. Diener, and M. Diener, "Le probleme de la “chasse au canard,” C. R. Acad. Sci. Paris (Ser. I) {\bf 286}, 1059 (1978).
\bibitem{canard2} E. Benoit, J.-L. Callot, F. Diener, and M. Diener, "Chasse au canards," Collect. Math. {\bf 32}, 37 (1981).
\bibitem{izhikevich} E. M. Izhikevich, "Neural excitability, spiking and bursting," Int. J. Bifurcation Chaos Appl. Sci. Eng. {\bf 10}, 1171 (2000).
\bibitem{berge} P. Berge, Y. Pomeau and C. Vidal {\it Order Within Chaos: Towards a Deterministic Approach to Turbulence} (Wiley, 1984).
\bibitem{yang} H. Yang and E. Ding, “Synchronization of chaotic systems and on-off intermittency,” Phys. Rev. E {\bf 54}, 1361 (1996).
\bibitem{zhao} L. Zhao, Y.-C. Lai, and C.-W. Shih, “Transition to intermittent chaotic synchronization,” Phys. Rev. E {\bf 72}, 036212 (2005).
\bibitem{choudhary} A. Choudhary, C. Mitra, V. Kohar, S. Sinha, and J. Kurths, “Small-world networks exhibit pronounced intermittent synchronization,” Chaos {\bf 27}, 111101 (2017).
\bibitem{vera} V. P. Vera-Ávila, J. R. Sevilla-Escoboza, and I. Leyva, "Complex networks exhibit intermittent synchronization," Chaos {\bf 30} 103119 (2020).
\bibitem{ahn}S. Ahn and L. L. Rubchinsky, “Potential mechanisms and functions of intermittent neural synchronization,” Front. Comput. Neurosci. {\bf 11}, 1–10 (2017).
\bibitem{park} C. Park, R. M. Worth, and L. L. Rubchinsky, “Fine temporal structure of beta oscillations synchronization in subthalamic nucleus in Parkinson’s disease,” J. Neurophysiol. {\bf 103}, 2707–2716 (2010).
\bibitem{rat} S. Ratnadurai-Giridharan, S. E. Zauber, R. M. Worth, T. Witt, S. Ahn, and L. L. Rubchinsky, “Temporal patterning of neural synchrony in the basal ganglia in Parkinson’s disease,” Clin. Neurophysiol. {\bf 127}, 1743–1745 (2016).
\bibitem{koro} A. A. Koronovskii, A. E. Hramov, V. V. Grubov, O. I. Moskalenko, E. Sitnikova, and A. N. Pavlov, “Coexistence of intermittencies in the neuronal network of the epileptic brain,” Phys. Rev. E {\bf 93}, 032220 (2016).
\bibitem{polonica} M. Ciszak, R. Meucci, "Spontaneous Transitions in Deterministic Networks," Acta Physica Polonica B {\bf 45}, 1157 (2014).
\end{thebibliography}
\end{document}